\documentclass[prl,aps,twocolumn,superscriptaddress]{revtex4}

\usepackage{epsfig}
\usepackage{natbib}
\usepackage{mathrsfs}
\usepackage{amsmath}

\begin{document}

\title{Experimental Approach to the Thermodynamics of the Pure Two-Dimensional
Spin-$1/2$ Triangular Lattice Antiferromagnet in Ba$_8$CoNb$_6$O$_{24}$}

\author{Y. Cui}
\thanks{These authors contributed equally to this study.}
\affiliation{Department of Physics and Beijing Key Laboratory of
Opto-electronic Functional Materials $\&$ Micro-nano Devices, Renmin
University of China, Beijing, 100872, China}

\author{J. Dai}
\thanks{These authors contributed equally to this study.}
\affiliation{Department of Physics and Beijing Key Laboratory of
Opto-electronic Functional Materials $\&$ Micro-nano Devices, Renmin
University of China, Beijing, 100872, China}

\author{P. Zhou}
\affiliation{Department of Physics and Beijing Key Laboratory of
Opto-electronic Functional Materials $\&$ Micro-nano Devices, Renmin
University of China, Beijing, 100872, China}

\author{P. S. Wang}
\affiliation{Department of Physics and Beijing Key Laboratory of
Opto-electronic Functional Materials $\&$ Micro-nano Devices, Renmin
University of China, Beijing, 100872, China}

\author{T. R. Li}
\affiliation{Department of Physics and Beijing Key Laboratory of
Opto-electronic Functional Materials $\&$ Micro-nano Devices, Renmin
University of China, Beijing, 100872, China}

\author{W. H. Song}
\affiliation{Department of Physics and Beijing Key Laboratory of
Opto-electronic Functional Materials $\&$ Micro-nano Devices, Renmin
University of China, Beijing, 100872, China}

\author{L. Ma}
\affiliation{High Magnetic Field Laboratory, Chinese Academy of Sciences,
Hefei 230031, China}

\author{Z. Zhang}
\affiliation{State Key Laboratory of Surface Physics and Laboratory of
Advanced Materials, Department of Physics, Fudan University, Shanghai
200433, China}

\author{S. Y. Li}
\affiliation{State Key Laboratory of Surface Physics and Laboratory of
Advanced Materials, Department of Physics, Fudan University, Shanghai
200433, China}
\affiliation{Collaborative Innovation Center of Advanced Microstructures,
Nanjing 210093, China}

\author{G. M. Luke}
\affiliation{Department of Physics and Astronomy, McMaster University,
Hamilton L8S 4M1, Canada}
\affiliation{Canadian Institute for Advanced Research, Toronto M5G 1Z8, Canada}

\author{B. Normand}
\affiliation{Laboratory for Neutron Scattering and Imaging, Paul Scherrer
Institute, CH-5232 Villigen PSI, Switzerland}

\author{T. Xiang}
\affiliation{Institute of Physics, Chinese Academy of Sciences, Beijing 100190,
China}
\affiliation{Collaborative Innovation Center of Quantum Matter, Beijing 100190,
China}

\author{W. Yu}
\email{wqyu\_phy@ruc.edu.cn}
\affiliation{Department of Physics and Beijing Key Laboratory of
Opto-electronic Functional Materials $\&$ Micro-nano Devices, Renmin
University of China, Beijing, 100872, China}
\affiliation{Department of Physics and Astronomy, Shanghai Jiaotong
University, Shanghai 200240, China and Collaborative Innovation Center
of Advanced Microstructures, Nanjing 210093, China}



\maketitle

{\bf Frustrated quantum magnets pose well-defined questions concerning
quantum fluctuation effects and the nature of the many-body wavefunction, which
challenge theory, numerics, experiment and materials synthesis. The $S = 1/2$
triangular-lattice antiferromagnet (TLAF) presents a case where classical
order is strongly suppressed by quantum fluctuations, leading to extensive
renormalization of physical properties at all energy scales. However, purely
two-dimensional (2D) models are difficult to realise in the 3D world and
their physics is controlled by the Mermin-Wagner theorem, which describes
the dominant effects of additional thermal fluctuations. Here we report the
magnetic properties Ba$_8$CoNb$_6$O$_{24}$, whose Co$^{2+}$ ions have an
effective spin 1/2 and construct a regular TLAF with very large interlayer
spacing. We find no magnetic ordering down to 0.028 K, strong low-energy
spin fluctuations in qualitative agreement with theoretical analysis and
a diverging correlation length, all indicating a Mermin-Wagner trend towards
zero-temperature ordering in this ideal 2D system.}

The Mermin-Wagner theorem \cite{Mermin_PRL__1966_17} describes the combined
effects of quantum and thermal fluctuations in the restricted phase space of
low-dimensional systems, and dictates that in 2D a continuous symmetry can be
broken, allowing a finite order parameter, only at exactly zero temperature.
In practice, most experimental systems are subject to a weak 3D coupling which
stabilizes their semiclassical order, and so examples of ``Mermin-Wagner
order,'' meaning incipient order as $T \rightarrow 0$, are rare.

\begin{figure*}
\includegraphics[width=16cm, height=10.7cm]{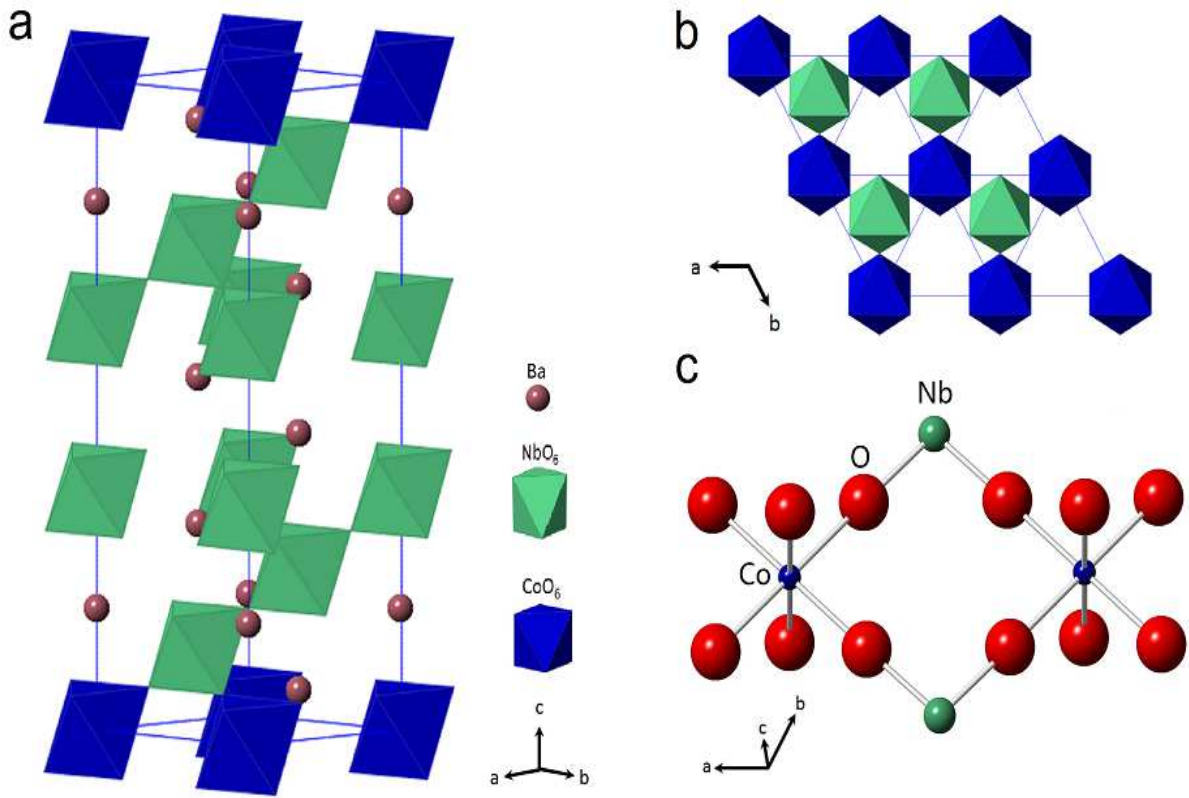}
\caption{\label{1} {\bf Lattice structure of Ba$_8$CoNb$_6$O$_{24}$.} {\bf a}
The triangular-lattice planes formed from the CoO$_6$ units are separated
by many nonmagnetic units, specifically 8 Ba$^{2+}$ and 6 NbO$_6$ layers.
{\bf b} A plane of CoO$_6$ octahedra, showing their triangular-lattice
configuration and the neighbouring, corner-sharing NbO$_6$ octahedra
which mediate the antiferromagnetic interactions. {\bf c} Local bonding
geometry of neighbouring Co$^{2+}$, Nb$^{5+}$ and O$^{2-}$ sites.}
\end{figure*}

\begin{figure*}
\includegraphics[width=16cm, height=12cm]{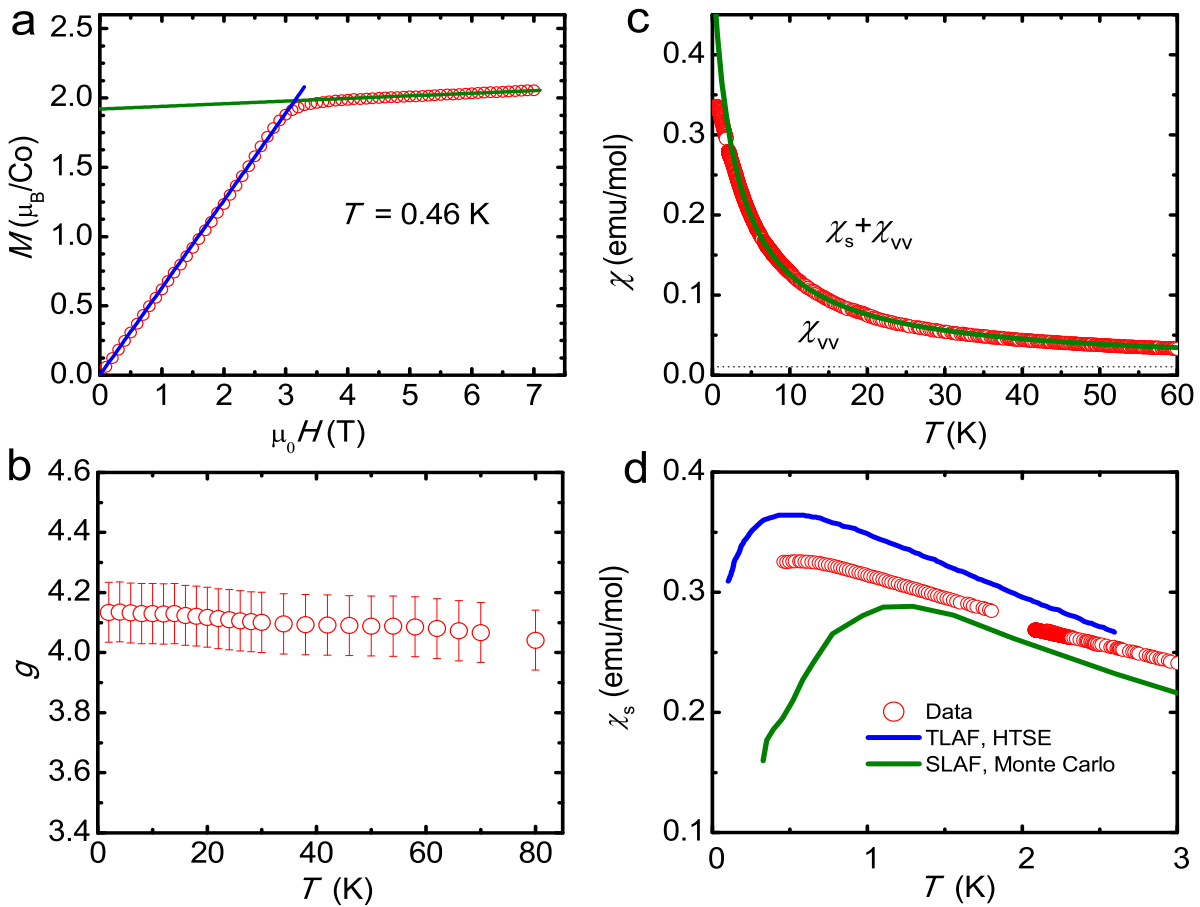}
\caption{\label{2} {\bf Magnetic properties of Ba$_8$CoNb$_6$O$_{24}$.}
{\bf a} Magnetization, $M$; blue and green straight lines represent linear-$H$
fits to the low- and high-field data, whose origins lie respectively in spin
and orbital contributions. {\bf b} $g$-factor as a function of temperature.
{\bf c} dc susceptibility measured under a field of 0.01 T. $\chi_{vv}$ is the
constant orbital contribution determined from $M$. The solid green line is
a fit to a Curie-Weiss form, with an offset of $\chi_{vv}$. {\bf d} Enlarged
view of the low-temperature spin susceptibility: the blue line shows HTSE
results for the Heisenberg TLAF, adapted from Ref.~\cite{Singh_PRL_1993_71};
the green line shows Monte Carlo results for the Heisenberg SLAF, adapted
from Ref.~\cite{Makivic_PRB__1991_43}. Both fits use the parameters
$J = 1.3$ K and $g = 4.13$.}
\end{figure*}

\begin{figure*}
\includegraphics[width=16cm, height=12cm]{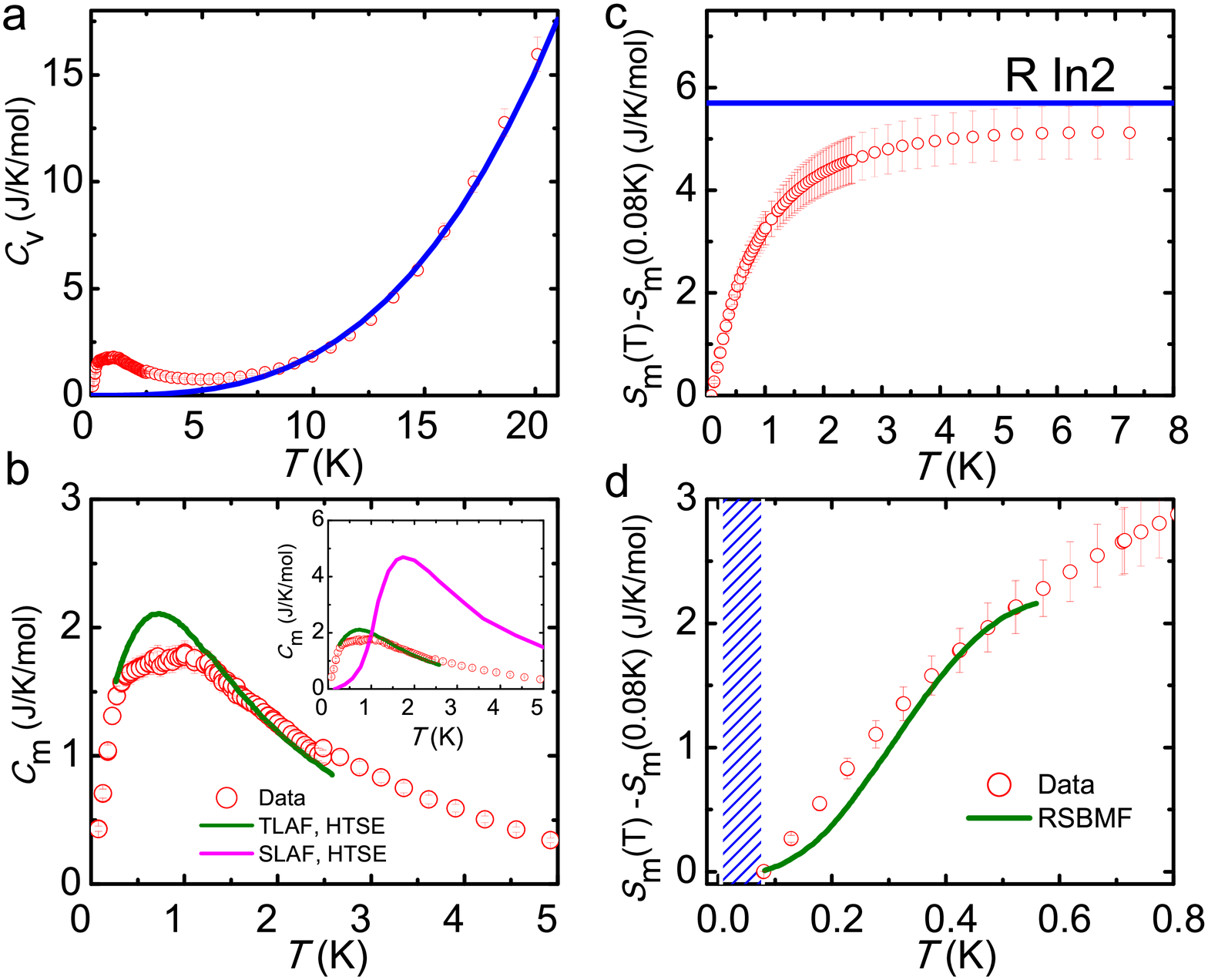}
\caption{\label{3} {\bf Specific heat and entropy analysis.} {\bf a} Specific
heat at zero field. The solid line is a fit to $C_v = a T^3$. {\bf b} Enlarged
view of the low-temperature magnetic specific heat, $C_m$, after subtracting
the phonon contribution. The green solid line is the HTSE result for the
Heisenberg TLAF, adapted from Ref.~\cite{Singh_PRL_1993_71} with $J = 1.3$ K.
Inset: TLAF $C_m(T)$ data from the main panel compared with the HTSE result
for the Heisenberg SLAF, adapted from Ref.~\cite{Bernu_PRB__2001_63}, for
the same value $J = 1.3$ K. {\bf c} Magnetic entropy, $S_m(T)$, obtained by
integrating the specific-heat data above 0.08 K. The solid line shows the
high-temperature limit, $S_m(\infty) = {\rm R} \ln(2S+1)$ in a spin-$S$
system, for $S = 1/2$. {\bf d} Enlarged view of $S_m(T)$ at low temperature,
showing the comparison with the RSBMF approach (see text), adapted from
Ref.~\cite{Singh_NJP_2012_14} with $J = 1.3$ K. Blue shading represents
the temperature region excluded from our analysis.}
\end{figure*}

The challenge of frustrated quantum magnetism is to characterize the effects
of quantum fluctuations in dimensions greater than 1. While the $S = 1/2$
square-lattice antiferromagnet (SLAF) with Heisenberg interactions has clear
magnetic order with a suppressed moment ($m_s \simeq 0.61 m_0$, where $m_0$
is the full moment for $S = 1/2$) and the kagome-lattice AF has no order at
all, the TLAF lies close to the boundary where the frustration is sufficient
to destroy $m_s$, leading to a range of exotic proposals \cite{Anderson_1973,
Momoi_JPSJ_1992,Chernyshev_PRL_2006_97,Starykh_RPP_2015_78}. Detailed studies
have demonstrated that the ground state of the Heisenberg TLAF does have a
finite semiclassical order, in a noncollinear $120^{\circ}$ structure
\cite{Bernu_PRL_1992_69,Capriotti_PRL_1999_82}, with a best estimate for
$m_s$ of $0.41 m_0$ \cite{White_PRL_2007}. However, the thermodynamic
properties of the TLAF have remained as a long-standing conundrum due to
the inadequacy of theoretical approximations, the limitations of numerical
approaches (including small system sizes in exact diagonalization, the
minus-sign problem in quantum Monte Carlo and the 1D restriction on
density-matrix renormalization-group methods) and the absence of pure 2D
systems for experimental investigation. A full understanding of the TLAF
would also aid the understanding of other novel quantum states, most
notably spin liquids \cite{Balents_nature_2010}, unconventional
superconductors \cite{Norman_2013} and systems with complex magnetic
order \cite{Yamamoto_PRL_2014_112}, in all of which geometric frustration
has an essential role.

A number of spin-1/2 TLAF compounds has been synthesized recently
\cite{Coldea_PRL__2001_86, Shimizu_PRL__2003_91}. Of most interest are
the materials Ba$_3$CoSb$_2$O$_9$ and Ba$_3$CoNb$_2$O$_9$, which have
perfectly regular lattices of Co$^{2+}$ ions and preservation of inversion
symmetry close to the plane, so that Dzyaloshinskii-Moriya (DM)
interactions are too small to break the continuous symmetry
\cite{Shirata_PRL_2012,Zhou_PRL_2012_109,Susuki_PRL__2013,Zhou_PRB_2014_89,
Yokota_PRB__2014_90}. It is found that these systems are close to a pure
Heisenberg interaction, albeit with a weak exchange anisotropy ($J_x = J_y
\neq J_z$). Magnetic order occurs at finite temperatures due to weak
interplane coupling, which makes the Mermin-Wagner theorem inapplicable.
The rich variety of competing magnetically ordered phases at finite applied
magnetic fields \cite{Yamamoto_PRL_2014_112} provides evidence of the
expected frustration effects. Available theoretical approaches
\cite{Singh_PRL_2006_12,Chubukov_PRL__1994_72, Chubukov_PRB__2006_74,
Mourigal_PRB__2013_88, Singh_NJP_2012_14} suggest the presence of weakly
dispersive excitations, which have not as yet been observed but whose effects
on the thermodynamic properties of the TLAF lie beyond the predictions of the
nonlinear-$\sigma$ model \cite{Singh_PRL_1993_71, Singh_PRB__2006_74,
Singh_NJP_2012_14}.

Here we report an experimental approach to the thermodynamic properties of
the purely 2D, spin-1/2 TLAF close to the Heisenberg point, which we make by
studying the dielectric compound Ba$_8$CoNb$_6$O$_{24}$. This system has
Co$^{2+}$ triangular layers separated by a very large interlayer spacing,
$c \simeq 18.9$ \r{A}. By combining magnetization, $g$-factor, susceptibility,
specific-heat and $^{93}$Nb nuclear quadrupole resonance (NQR) measurements down
to temperatures of 0.028 K, we reveal that i) the system has an effective
spin-$1/2$ with a nearest-neighbour antiferromagnetic exchange coupling
$J \simeq 1.3$ K, ii) there is no spontaneous magnetization down to 0.028 K
under zero field but a steeply increasing correlation length upon cooling
and iii) the low-temperature susceptibility and specific heat provide a
qualitative validation of high-temperature series-expansion (HTSE) and
reconstructed Schwinger-boson mean-field (RSBMF) approaches to the TLAF,
but quantitative differences remain at the lowest temperatures.
(ii) may be the first experimental illustration of the textbook Mermin-Wagner
theorem concerning zero-temperature magnetic ordering in a 2D system. (iii)
shows anomalously low energy scales and discrepancies from theory which lie
in sharp contrast to the properties of the SLAF. Thus Ba$_8$CoNb$_6$O$_{24}$
constitutes a model system for characterizing the interplay of geometrical
frustration, quantum and thermal fluctuations.

\bigskip
\noindent
{\bf Material}
\medskip

Polycrystalline Ba$_8$CoNb$_6$O$_{24}$ samples were synthesized by a solid-state
reaction method \cite{Mallinson_Angew_2005_117}. The material crystallizes
in the space group $P\bar3m1$, illustrated in Fig.~\ref{1}a. Co$^{2+}$ ions
in corner-shared CoO$_6$ octahedra construct perfect layers of regular
triangular lattices (Fig.~\ref{1}b), with neighbouring Co$^{2+}$ planes
separated by eight Ba$^{2+}$ and six NbO$_6$ layers. The lattice parameters
\cite{Mallinson_Angew_2005_117} are $a = 5.789813$ \r{A}, almost identical
to that of Ba$_3$CoNb$_2$O$_9$ ($a = 5.7737$ \r{A}), but $c = 18.89355$ \r{A},
which is approximately three times longer.

It is necessary first to establish the effective spin of the Co$^{2+}$ ions in
Ba$_8$CoNb$_6$O$_{24}$. Figure \ref{2}a shows the magnetization, $M$, as a
function of field at a fixed temperature $T = 0.46$ K. $M$ increases rapidly
and linearly up to a field $H_S = 3.12 \pm 0.04$ T, beyond which a weak
linear increase is also observable (the Van Vleck orbital contribution
\cite{Zhou_PRB_2014_89}, $\chi_{vv} \simeq 0.019\mu_B$/Co/T, or 0.0106 emu/mol).
$H_S$ is the saturation field required to polarize fully the magnetic moment
of Co$^{2+}$, as also measured for Ba$_3$CoSb$_2$O$_9$ \cite{Shirata_PRL_2012,
Zhou_PRL_2012_109,Susuki_PRL__2013} and Ba$_3$CoNb$_2$O$_9$
\cite{Zhou_PRB_2014_89,Yokota_PRB__2014_90}. The saturation moment of
the Co$^{2+}$ ions deduced from Fig.~\ref{2}a is $m_s = 1.92 \pm 0.1 \mu_B$.
Measurements of the $g$-factor at different temperatures, shown in
Fig.~\ref{2}b, approach a constant value $g = 4.13 \pm 0.1$ below 20 K.
The effective spin is therefore $m_s/g\mu_B \simeq 0.464$, which suggests
a spin-$1/2$ state for Co$^{2+}$. This is consistent with the crystal-field
analysis for Co$^{2+}$ in an octahedral environment \cite{Shirata_PRL_2012,
Shiba_JPSJ_2003_72}, where a Kramers doublet is formed due to strong
spin-orbit coupling, leading to an effective spin-$1/2$ with a large
$g$-factor at low temperatures.

The magnetic exchange coupling can be estimated from the same data.
From the corner-sharing geometry of neighbouring CoO$_6$ and NbO$_6$ octahedra,
the dominant magnetic interactions between in-plane Co$^{2+}$ spins occur by
Co$^{2+}$-O$^{2-}$-O$^{2-}$-Co$^{2+}$ and Co$^{2+}$-O$^{2-}$-Nb$^{5+}$-O$^{2-}$-Co$^{2+}$
superexchange couplings (Fig.~\ref{1}c \cite{Zhou_PRB_2014_89}); the very long
paths make the interaction strength extremely sensitive to geometrical details
and preclude all but nearest-neighbour couplings. For the effective $S = 1/2$
Co$^{2+}$ ions one expects an XXZ spin model of the form $H = \sum_{\langle ij
\rangle} J_x (S_i^x S_j^x + S_i^y S_j^y) + J_zS_i^z S_j^z$, where $\langle ij
\rangle$ denotes nearest-neighbour Co$^{2+}$ spins \cite{Shirata_PRL_2012,
Zhou_PRL_2012_109,Susuki_PRL__2013,Zhou_PRB_2014_89,Yokota_PRB__2014_90}.
In Ba$_3$CoSb$_2$O$_9$, the exchange coupling is close to isotropic
(Heisenberg), with $J_x = J_y \approx J_z \approx$ 18.2 K
\cite{Shirata_PRL_2012,Zhou_PRL_2012_109,Susuki_PRL__2013}. In
Ba$_3$CoNb$_2$O$_9$, double magnetic transitions at 1.36 K and 1.10 K
in zero field are consistent with a weak easy-axis anisotropy, $J_z > J_x$
\cite{Zhou_PRB_2014_89,Yokota_PRB__2014_90,Matsubara_JPS__1982_51,
Miyashita_JPSJ_1985_54}. Because Ba$_8$CoNb$_6$O$_{24}$ and Ba$_3$CoNb$_2$O$_{9}$
have an almost identical planar structure, very similar exchange couplings
are expected: using the result $4.5 J = g \mu_B H_S$ for the Heisenberg TLAF
at $T = 0$ \cite{Yamamoto_PRL_2014_112}, from Fig.~\ref{2}a we estimate
that $J = 1.3 \pm 0.1$ K for Ba$_8$CoNb$_6$O$_{24}$. We comment that this
procedure may underestimate $J$ by a small amount, because our measurement
of $H_S$ is not made at zero temperature, but this effect is included within
the error bar. In the absence of single crystals, we do not speculate on the
anisotropy of the exchange couplings in Ba$_8$CoNb$_6$O$_{24}$, but from the
properties of Ba$_3$CoNb$_2$O$_9$ and Ba$_3$CoSb$_2$O$_9$ one expects to be
close to the Heisenberg model ($J_x \simeq J_z$).

\bigskip
\noindent
{\bf Thermodynamic Properties}
\medskip

Turning to thermodynamic properties, the dc susceptibility, $\chi(T)$, of
the sample, measured under a field of 0.01 T, is shown in Fig.~\ref{2}c.
From 20 K down to 1 K, $\chi(T)$ increases upon cooling and can be separated
into the two contributions $\chi(T) = \chi_{vv} + \chi_{s}(T)$. $\chi_{s}(T)$
follows a Curie-Weiss (CW) form, $\chi_{s}(T) = a/(T + \theta)$, which
represents the average contribution of the coupled Co$^{2+}$ spins and
returns a Weiss constant $\theta \approx 3.5 \pm 0.5$ K. At low temperatures,
$\chi_s(T)$ is expected to fall below the CW form when $T < J$ and the spins
become correlated (Fig.~\ref{2}c). For comparison, in Fig.~\ref{2}d we show
$\chi_s(T)$ obtained by HTSE on the Heisenberg TLAF, adapted from
Ref.~\cite{Singh_PRL_1993_71} by using the parameters $J = 1.3$ K and
$g = 4.13$ deduced from $M$ (Fig.~\ref{2}a).

\begin{figure*}
\includegraphics[width=16cm, height=6cm]{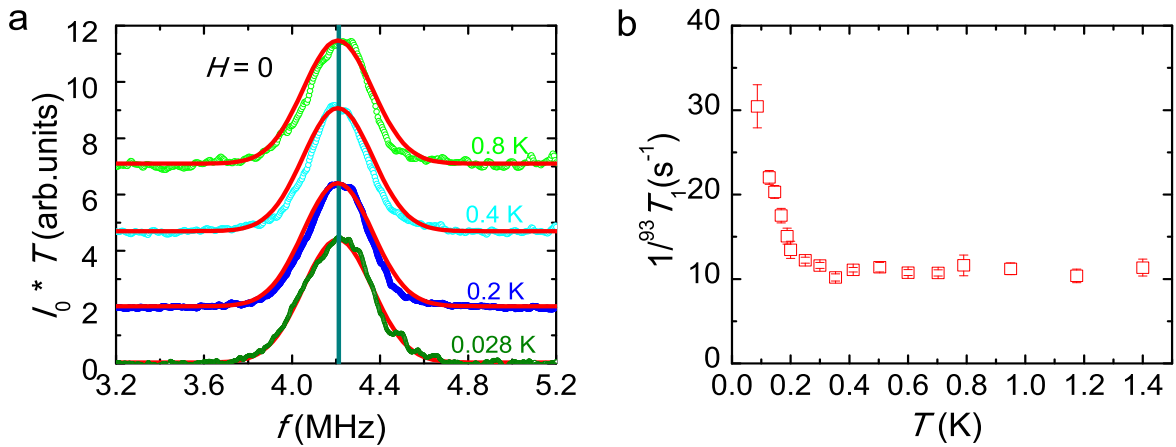}
\caption{\label{4} {\bf NQR measurements of Ba$_8$CoNb$_6$O$_{24}$.}
{\bf a} Zero-field $^{93}$Nb NQR spectra for temperatures from 0.8 K to 0.028
K, with the measured intensity multiplied by $T$. Red lines represent the
Gaussian fit to the spectrum at 0.028 K overlaid on all four datasets.
{\bf b} Spin-lattice relaxation rate, $1/^{93}T_1$, as a function of
temperature. The upturn below 0.2 K indicates a progressive increase of
the correlation length upon cooling.}
\end{figure*}

At the qualitative level, the HTSE fit is completely consistent with our
data, in particular the fact that the peak in $\chi_s(T)$ occurs at the
anomalously low temperature $T \approx 0.4 J$; this can be contrasted
with the behaviour of the unfrustrated SLAF (Fig.~\ref{2}d
\cite{Makivic_PRB__1991_43}), where the peak appears at $T \approx J$.
Quantitatively, the HTSE fit lies above our data by approximately 10\%,
and we comment that a better fit would be obtained with a value $J \simeq
1.6$ K. However, such a value would make it impossible to fit $M$
(Fig.~\ref{2}a) without driving $g$ (Fig.~\ref{2}b) well outside the error
bars on our measurement. Because $M$ constitutes the most direct and accurate
access to $J$ in experiment, we continue to be guided by the value it yields.
We comment also that the HTSE is by nature an approach from high temperatures,
which reaches its limits at the anomalously low temperatures of the $\chi_s$
peak in the TLAF, and its use requires careful choice of representative
Pad\'e approximants.

Further valuable thermodynamic information is provided by the specific heat,
$C_v(T)$, shown in Fig.~\ref{3}a. The absence of any peak or cusp feature
in $C_v$ suggests that magnetic ordering is absent to the lowest temperature
we can access. Because $C_v$ falls rapidly from 20 K to 7.5 K, following an
exact $T^3$ behaviour, we use this to subtract the presumed phonon contribution
and isolate the magnetic specific heat, $C_m(T)$. This procedure can be
followed with a high degree of confidence because the characteristic
energy scales of the phonons are manifestly very high compared to the
magnon contributions in this system, which peak at 1 K. However, we caution
that even very small residual uncertainties may be important in the entropy
analysis below, and contribute to the error bars we display. $C_m(T)$, shown
in Fig.~\ref{3}b, confirms the absence of ordering features, is dominated by
a broad peak at $T \approx 1.0$ K and below 0.3 K falls rapidly towards
$C_m = 0$ with no evidence for an actived form (i.e.~for a spin gap). Once
again we compare our data with the HTSE result \cite{Singh_PRL_1993_71} for
the Heisenberg TLAF with $J = 1.3$ K, finding a semi-quantitative level of
agreement over the available range of the HTSE data (0.3 K $\le T \le$ 2.5 K).
As for the susceptibility, the peak in $C_m(T)$ lies at a value anomalously
low compared with the energy scale of the SLAF, as deduced from QMC
\cite{Makivic_PRB__1991_43} and HTSE methods \cite{Wang_PRB_1991_44,
Bernu_PRB__2001_63} and shown in the inset of Fig.~\ref{3}b; this result
reflects directly the effects of frustration in suppressing the overall
energy scale of the magnetic excitations \cite{Singh_PRB__2006_74}.

The magnetic entropy, $S_m(T) = \int C_m/T dT$, which we calculate from
our $C_m(T)$ data by integrating above 0.08 K, is shown in Fig.~\ref{3}c.
At 7.5 K, we estimate that $S_m(T)$ saturates $90 \pm 10$\% of its
total value, $S_m(\infty) = R \ln 2$ for a spin-1/2 system. Because
of the errors accumulated in the integration, including those from the
phonon subtraction, we cannot draw any meaningful conclusions about
possible missing entropy from this result, other than that it is small.
At low temperatures, however, we find that $S_m(T)$ has a very rapid
initial increase from $T = 0.08$ K, with 33$\%$ of $S_m(\infty)$ recovered
by $T = 0.3J$ (Fig.~\ref{3}d). Here we compare our data to the RSBMF approach
\cite{Singh_NJP_2012_14}, a modified Schwinger-boson technique which counts
the correct number of physical states and thus is expected to capture the key
properties of the TLAF in the low-temperature regime. By comparison with our
data, the RSBMF formalism provides quantitatively accurate state-counting for
$T > 0.3 J$, which we note is near the limit of the authors' claims for its
validity. However, the form of $S_m(T)$ at lower temperatures is not well
described, a result on which we comment briefly below. Nevertheless, we
note for perspective that the type of nonlinear-$\sigma$-model approach so
effective for the SLAF recovers only 5$\%$ of the total entropy at $T = 0.3
J$ in the TLAF \cite{Singh_PRL_1993_71,Chubukov_JPCM_1994_6} and therefore
appears incapable of providing a suitable account of frustrated systems.

\bigskip
\noindent
{\bf NQR Measurements}
\medskip

To probe both static and low-energy magnetic properties, we present
low-temperature $^{93}$Nb NQR data. Here we report only the signal with the
shortest spin-lattice relaxation time, $^{93}T_1$, which is over three orders
of magnitude faster than the other times present and arises in all probability
from the Nb sites closest to the Co$^{2+}$ layers (i.e.~with strong hyperfine
coupling to the Co$^{2+}$ moments). The $^{93}$Nb ($I = 9/2$) NQR spectra for
excitations between $I_z = \pm 9/2$ and $I_z = \pm 7/2$ are shown in
Fig.~\ref{4}a, for temperatures from 0.8 K to 0.028 K and with intensity
normalized by the Zeeman factor, $1/T$. At $T = 0.028$ K, the spectrum is
well fitted by a Gaussian function (red line), and this fit is applied also
to the spectra at 0.2 K, 0.4 K and 0.8 K. Clearly all spectra are centred
at the same frequency, $f = 4.212$ MHz, and the echo intensity remains
constant with temperature. The absence of any signal loss and of any NQR
frequency shift completely excludes any magnetic order down to 0.028 K,
consistent with the specific-heat data. Because the $^{93}$Nb nucleus is
located directly above the centre of one triangle composed of three
Co$^{2+}$ ions (Fig.~\ref{1}b), the static hyperfine field on $^{93}$Nb
should be finite with an off-diagonal component, if the Co$^{2+}$ moments
order with the $120^{\circ}$ coplanar pattern \cite{Yamamoto_PRL_2014_112},
and thus it is highly unlikely that magnetic order could be missed by the
NQR spectra.

The NQR spin-lattice relaxation rate, $1/^{93}T_1$, shown in Fig.~\ref{4}b,
probes the low-energy spin fluctuations. In general, $1/T_1 = \sum_q A_{\rm hf}
(q) \, {\rm Im} \chi^{\pm} (q,\omega)/\omega|_{\omega\rightarrow 0}$, where $A_{\rm
hf}$ is the hyperfine coupling constant and $\chi^{\pm}(q,\omega)$ the
transverse dynamic susceptibility. The fact that $1/^{93}T_1$ is of order
$10$ s$^{-1}$ indicates very strong hyperfine coupling between $^{93}$Nb and
the Co$^{2+}$ spins. $1/^{93}T_1$ also contains no evidence for long-range
order, although its moderate upturn below 0.2 K indicates an increasing
correlation length, precisely as would be expected if the system approaches
the zero-temperature magnetic order anticipated by the Mermin-Wagner theorem
\cite{Mermin_PRL__1966_17}. To our knowledge, this is the first direct
observation of incipient ``Mermin-Wagner'' order in a purely 2D magnetic
system.

\bigskip
\noindent
{\bf Discussion}
\medskip

For a deeper understanding of the anomalous low energy scale observed in the
thermodynamic quantities of Fig.~\ref{3}, we turn to the most sophisticated
HTSE \cite{Singh_PRL_2006_12,Singh_PRB__2006_74}, self-consistent spin-wave
\cite{Chubukov_PRB__2006_74,Mourigal_PRB__2013_88} and RSBMF treatments
\cite{Singh_NJP_2012_14} of the Heisenberg TLAF. These demonstrate that
the magnetic excitation band is both extremely flat and unusually low-lying,
with strong weight around $E \approx 0.6J$. This is a direct consequence of
the strong frustration of the TLAF and stands in sharp contrast to the SLAF,
where the bands disperse uniformly up to energies $E \approx 2J$
\cite{Jarrell_PRL_1992_68}. The frustration-renormalized energy scale is
completely consistent with the temperature-dependence we benchmark in the
peak features of $\chi(T)$ and $C_m(T)$. These unusual low-energy excitations
of the TLAF have been variously described as ``roton-like'' or as evidence of
fermion deconfinement \cite{Singh_PRL_2006_12,Chubukov_PRB__2006_74}.

To our knowledge, Ba$_8$CoNb$_6$O$_{24}$ allows the first direct comparison
between theory and a real material for the spin-1/2 Heisenberg TLAF. Small
but finite discrepancies do appear at the quantitative level. The HTSE
results for $C_m(T)$ (Fig.~\ref{3}b) appear to overestimate the peak
contributions in our data and underestimate those at higher energies. The
RSBMF approach appears to underestimate the low-temperature entropy. Both
results may be explained if the ``roton gap,'' the effective bandwidth or
other features affecting the density of states in the spin spectrum, all of
which require a correct accounting for quantum fluctuation effects, are not
reproduced perfectly by the theoretical or numerical approaches applied.
However, we caution that the mismatch between our data and the theoretical
results shown in Figs.~\ref{2} and \ref{3} cannot be interpreted unambiguously
as evidence for shortcomings in the theories, as it may be caused by small
additional terms in the magnetic Hamiltonian, most notably a weakly
non-Heisenberg anisotropy in the exchange couplings. We defer a further
analysis of this point until single crystals become available.

In summary, we have performed experimental measurements of the thermodynamic
and NQR response of a purely 2D TLAF. This model system is realized in the
compound Ba$_8$CoNb$_6$O$_{24}$, where the very large separation of magnetic
layers precludes any 3D coupling. The material illustrates clearly the
Mermin-Wagner theorem for a 2D system by the absence of the magnetic
ordering down to 0.028 K but a possible zero-temperature order reflected
in the increase of $1/T_1$ below 0.2 K. The anomalously low thermodynamic
energy scales gauge the strong suppression and flattening of the excitation
bands and our measurements benchmark the efficacy of advanced theoretical
and numerical approaches in reproducing these frustration-induced properties
of the TLAF.

\bigskip
\noindent
{\bf Methods}
\medskip

Polycrystals of Ba$_8$CoNb$_6$O$_{24}$ were synthesized by a solid-state
reaction method \cite{Mallinson_Angew_2005_117}. Magnetization and
susceptibility data were measured in a PPMS-VSM for temperatures $T > 2$ K
and in a $^3$He SQUID system for $0.6 K < T < 1.8$ K. The temperature-dependent
$g$-factors were obtained by field-sweep ESR at a fixed frequency 9.397 GHz.
The specific heat was measured in a PPMS dilution refrigerator (DR), which
reached temperatures down to 0.08 K. The $^{93}$Nb ($I = 9/2$) NQR signal was
detected by the spin-echo technique in a DR system reaching temperatures down
to 0.028 K. The spin-lattice relaxation rate, $1/^{93}T_1$, was determined by
the inversion-recovery method, and the spin recovery fitted by the standard
function $I(t) = I(\infty) - a[4/33 e^{-3t/T_1} + 80/143 e^{-10t/T_1} + 49/165
e^{-21t/T_1} + 16/715 e^{-36t/T_1}]$ for $I$ = 9/2 spins
\cite{MacLaughlin_PRB__1971_4}. The theoretical values for the thermodynamic
quantities $\chi(T)$, $C_m(T)$ and $S_m(T)$ were digitized from the cited
literature and scaled appropriately.

\bigskip
\noindent
{\bf Acknowledgements}

\noindent
We thank A. Honecker, J. Richter, R. Yu and Y. Zhou for helpful discussions.
This Work was supported by the National Science Foundation of China (Grant
Nos.~11222433 and 11374364), the Ministry of Science and Technology of China
(Grant Nos.~2016YFA0300503 and 2016YFA0300504) and the Fundamental Research
Funds for the Central Universities and the Research Funds of Renmin University
of China (Grant No.~14XNLF08).

\bigskip
\noindent
{\bf Author contributions}

\noindent
The project was conceived by W.Q.Y.
The crystals were grown by J.D.
High-temperature ($T > 2.5$ K) susceptibility and specific-heat measurements
were made by J.D.
DR NQR measurements were performed by Y.C., with assistance from P.Z., P.S.W.,
T.R.L. and W.H.S.
Z.Z. and S.Y.L. made DR specific-heat measurements.
L.M. made $g$-factor measurements.
Susceptibility and magnetization measurements in the $^3$He-SQUID were made
by G.M.L.
The theoretical framework was provided by B.N.~and T.X.
Data refinement and figure preparation were performed by Y.C.~and W.Q.Y.
The text was written by B.N.~and W.Q.Y.

\bigskip
\noindent
{\bf Additional information}

\noindent
The authors declare no competing financial interests. Correspondence and
requests for materials should be addressed to W.Q.Y. (wqyu\_phy@ruc.edu.cn).

\end{document}